\documentstyle[11pt,newpasp,twoside,epsf]{article}
\def\edcomment#1{\iffalse\marginpar{\raggedright\sl#1\/}\else\relax\fi}
\reversemarginpar

\begin{document}
\title{A wide field survey of the distant rich cluster Cl0024+1654}
\author{T. Treu, R.S. Ellis, P. Trivedi}
\affil{California Institute of Technology, MS 105-24, Pasadena, CA 91125}
\author{J.-P. Kneib}
\affil{Observatoire Midi-Pyr\'en\'ees, UMR5572, 14 Av. \'Edouard Belin, 31400 Toulouse, France}
\author{A. Dressler, A. Oemler}
\affil{Carnegie Observatories, 813 Santa Barbara Street, Pasadena, CA 91101}
\author{P. Natarajan}
\affil{Department of Astronomy, Yale University, P.O. Box 208101, New Haven, CT 06250}
\author{I.~R. Smail}
\affil{Department of Physics, University of Durham, South Road, Durham DH1 3LE, UK}

\begin{abstract}
We describe the first results from a comprehensive study of the
distant cluster Cl0024+1654 ($z=0.39$) based upon a pattern of 38
mosaiced HST-WFPC2 images extending to radii $\sim 5$ Mpc. These are
being analysed in conjunction with extensive spectroscopy conducted
with the CFHT, WHT, and Keck Telescopes.  The overall goal is to
understand the morphological transformations and associated short-term
star formation histories of representative numbers of infalling field
galaxies in the context of the cluster potential as defined by weak
lensing studies. Our HST database contains over 2000 galaxy
morphologies to $I=22.5$. Spectroscopic data and HST morphologies are
currently available for about 215 members over an unprecedented range
of environments. We confirm the existence of a well-defined
morphology-density relation over a large dynamic range within a single
system at a significant look-back time. Tentative trends in the E/S0
fraction as a function of radius are discussed. A weak lensing signal
in the background galaxies has been detected at the cluster periphery
and its inversion demonstrates only marginal substructure. A
statistically-significant galaxy-galaxy lensing signal has also been
seen for the cluster members. Further work will relate radial
dependencies in the dark matter and halo masses in the context of
spectroscopic and morphological diagnostics of truncated star
formation.

\end{abstract}

\section{Introduction}

The ability of the Hubble Space Telescope (HST) to resolve galaxy
morphologies at cosmologically-significant look-back times has enabled
great progress in understanding the origin of the morphology density
relation and, {\em inter alia}, the influence of the environment on
galaxy morphology. On the basis of HST images of several clusters at
$z\sim0.3-0.5$, the Morphs team (Dressler et al.\ 1997) determined a
remarkable decline with redshift in the fraction of S0s with a
corresponding rise in that for spirals. They deduced that spirals were
transformed to S0 galaxies remarkably recently, most likely because
their gas was removed by processes such as tidal effects and ram
pressure stripping.

Alongside this evolutionary signal, strong evolution is also found in
star formation and morphological characteristics of field galaxies
(Glazebrook et al.\ 1995; Lilly et al.\ 1996). Understanding the role
that gas-rich field galaxies play in fueling the transformations
inferred in cluster cores remains an interesting issue. As field
galaxies most likely fall into clusters at all epochs, the origin of
the recent demise of spirals is puzzling.

A promising way forward is to study, in considerable detail, the
transformations occurring {\em in situ} in a single system from
the virialised core to the periphery where field galaxies infall
on radial orbits. This is preferable to drawing deductions based
on studies of many clusters, observed at different redshifts, each
at an unknown stage in its evolutionary history.

Because of the small field of view of WFPC-2, most morphological
studies have been limited to the very central regions of distant
clusters (e.g. Smail et al.\ 1997), and correspondingly little is
known about the properties of the infalling galaxy population at large
radii. Ground based studies (Abraham et al.\ 1996a) have explored the
{\em integrated} properties of galaxies to large cluster radii,
finding evidence for radial trends in diagnostics of recent star
formation. It seems likely that star formation is suppressed as
galaxies infall, although the physical mechanisms and timescales are
still quite controversial (Poggianti et al.\ 1999; Balogh et al.\
2000). These latter uncertainties may be overcome if the timescales of
stellar activity/truncation can be linked to those determined
dynamically in a known gravitational potential (e.~g. from weak
lensing signal).

We present the first results from a wide field HST survey of the rich
cluster Cl0024+1654 (hereafter 0024; $z=0.39$) whose goal is to
address many of the above questions. In the following, the Hubble
constant is H$_0=50h_{50}$kms$^{-1}$ Mpc$^{-1}$, assuming $h_{50}=1.3$
when necessary. The matter density of the Universe and the
cosmological constant are $\Omega=0.3$ and $\Omega_{\Lambda}=0.7$
respectively.

\section{Overview of Database}

0024 ($z=0.39$) is an ideal cluster with which to begin a detailed
dynamical, morphological and stellar population study of this
type. The results can be contrasted with similar pioneering
ground-based studies of the Coma cluster by Kent \& Gunn (1982) and
Colless \& Dunn (1996) and of Abell 2390 ($z=0.23$) by Abraham et al.\
(1996a). The distance to 0024 is sufficient that only a relatively
modest number (38) of WFPC2 images is needed for a well-sampled study
almost to the turn-around radius. The cluster appears regular and
symmetric in both its galaxy distribution and X-ray morphology. The
availability of over 600 spectroscopic redshifts from the CFHT/WHT
(Czoske et al.\ 2001a) provides a major starting point for a
comprehensive dynamical study. As is often the case with such
extensive sampling of the velocity field (c.f. Coma), complications
emerge. The redshift distribution of confirmed members already
suggested the existence a foreground system of modest mass. However,
recent work by Czoske et al.\ (2001b) suggests the dynamical
distribution of the innermost galaxies is consistent, although not
uniquely so, with a line-of-sight merger of two systems. In addition,
the central mass distribution can be independently reconstructed from
the quintuple gravitational arc system (Broadhurst et al.\ 2000).

The first component of our dataset is the wide-field WFPC2 data,
providing galaxy morphology and the weak shear to $\sim15$'($\sim
5$Mpc) from the cluster center. The primary imaging was accomplished
using 2 orbit exposures using the F814W filter, with parallel images
taken using STIS in clear CCD mode. As of July 2001 when the
conference was held, only 22/38 planned pointings had been
observed. However, the remainder arrived soon after and this report
presents the first preliminary analysis based on the entire dataset
(see Fig.~1).

\begin{figure}
\plotone{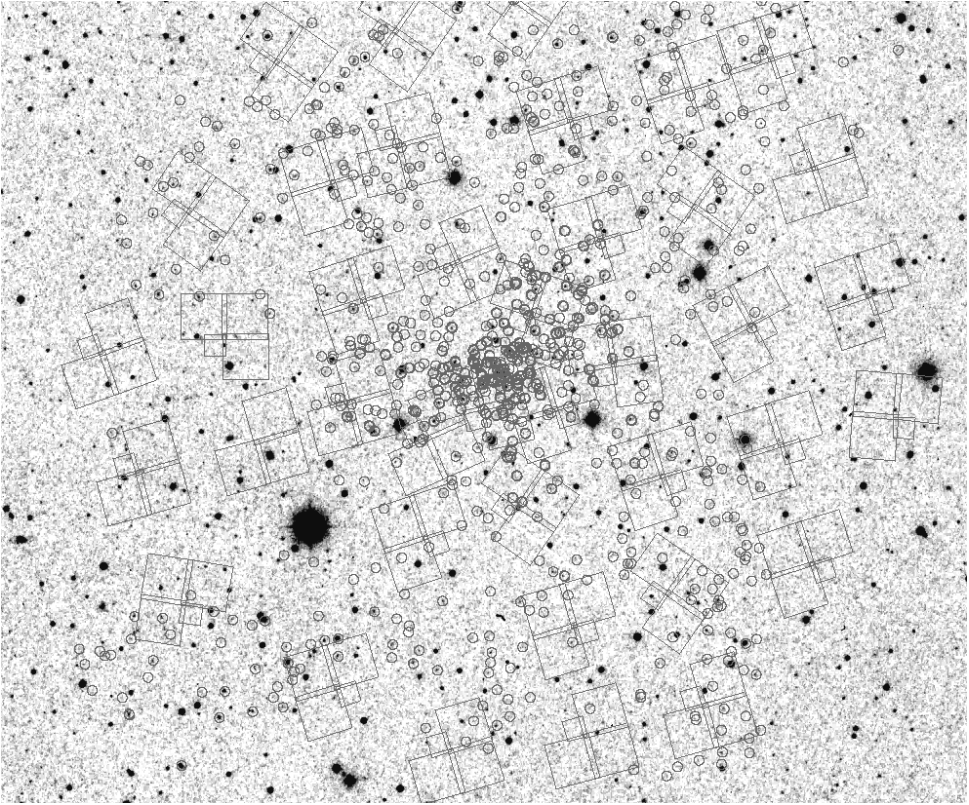} 
\caption{WFPC2 pointings and available redshifts (open circles).  The
working database of cluster 0024 ($z=0.39$) contains $\sim2000$ HST
morphologies and $\sim750$ spectroscopic redshifts.}
\end{figure}

Via comparisons with both the Morphs and Medium Deep Survey data taken
in the same F814W filter, it was determined that morphological
classifications in 0024 could be reliably performed within the MDS
classification system (Glazebrook et al.\ 1995, Abraham et al.\ 1996b)
to a limiting magnitude of $I=22.5$. The slightly deeper Morphs
exposure in the cluster core permits a similar classification to
$I=23$ (Smail et al.\ 1997). Systematic classifications performed by
one of us (RSE) yields a sample of 1800 objects ($\sim 600$ E/S0
$\sim800$ S, $\sim$ 400 Irr ), in addition to $\sim250$ already
classified in the core. Independent classification is currently being
performed by a second collaborator (AD).

The second component of our survey is the redshift survey. In the
outer fields ($>$1-2 Mpc from the center), the galaxy excess above
background per WFPC2 field to $I\sim22.5$ is within the observed
fluctuations in the field counts (Abraham et al.\ 1996a); to identify
infalling members, spectra are crucial. So far, we have collated 718
redshifts (Dressler et al.\ 1999; Czoske et al.\ 2001a; Owen 2001;
Metevier \& Koo 2001) including 334 members. Further spectroscopy in
the periphery is necessary because of the incomplete sampling by
earlier studies.

The third component is a spectroscopic campaign with the Keck
Telescopes, following up known members to $I=21$ and extending the
completeness of the redshift surveys to $I=23$. These data represent a
crucial part of the study of the infalling population providing
diagnostics of recent star formation and dynamical masses either
through stellar velocity dispersions or resolved emission line
rotation curves. A first attack in Oct 2001 yielded spectra for $\sim
100$ objects.

\section{First results}

\subsection{Galaxy population}

Our dataset provides the first comprehensive and uniform set of
morphological data across the range of environments from the
cluster core out to $\sim5$ Mpc ($\simeq$ ``field"). This has
enabled us to examine the suggestion that there are strong radial
gradients in the morphological mix and, in future, to tie these to
spectroscopic diagnostics which can be used to understand the
interplay between dynamical infall and the truncation of star
formation.

Whereas photometric techniques are effective in addressing the
statistical properties of background-subtracted galaxies in low
densities (Kodama et al.\ 2001), we seek here to identify members
individually so the dynamical and spectral diagnostics can be brought
into play. As an illustration of the difficulties associated with
statistical subtraction techniques, we show in Fig.~2, the run of
morphological type with radius from the cluster core to regions of
very low density. The left panel shows the $I<22.5$ galaxy density as
a function of cluster radius divided by broad morphological type. As
expected, E/S0 are more centrally concentrated than spirals, while
Irr/Unclass/Mergers show little evidence of any concentration. At
radii $\ga2$ Mpc, the observed density lies within the random
fluctuations of the average field density (dashed line; Abraham et
al.\ 1996b), and the background contamination can no longer be
reliably removed statistically. We can obtain a smoother
morphology-radius relation by averaging azimuthally, as shown in the
right panel of Fig.~2. The peaked distribution of E/S0s is now more
prominent and an excess in both E/S0s and spirals is seen to the
largest radius probed. However, this excess is critically dependent on
the assumed background count and its morphological
distribution. Astonishingly, the morphological counts (Glazebrook et
al.\ 1995, Driver et al. 1995) remain uncertain simply by virtue of
the limited area sampled by HST. This emphasises the importance of a
redshift survey (to a depth equivalent to that of local
determinations) to construct a reliable morphology-density relation.

\begin{figure}
\plottwo{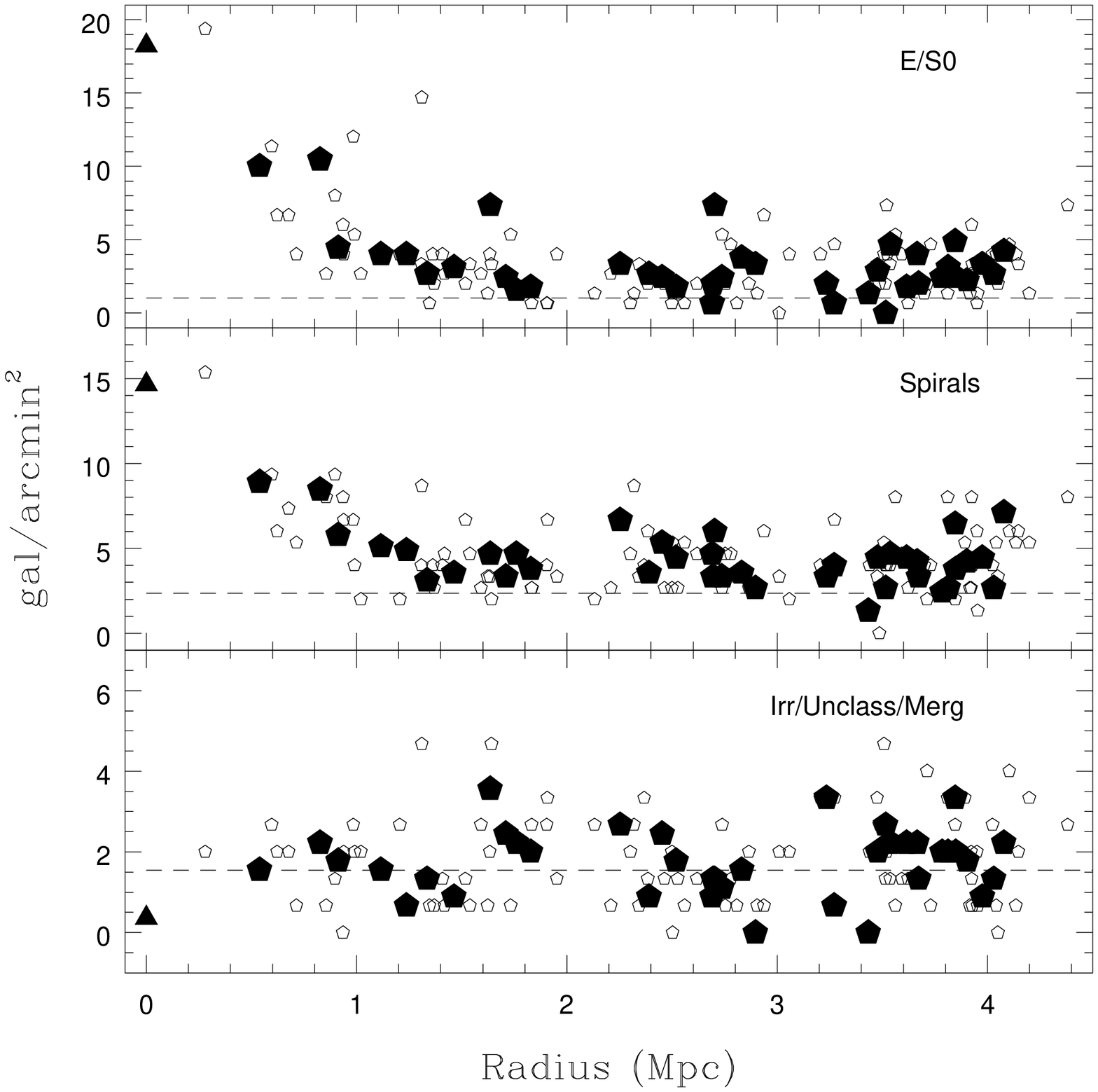}{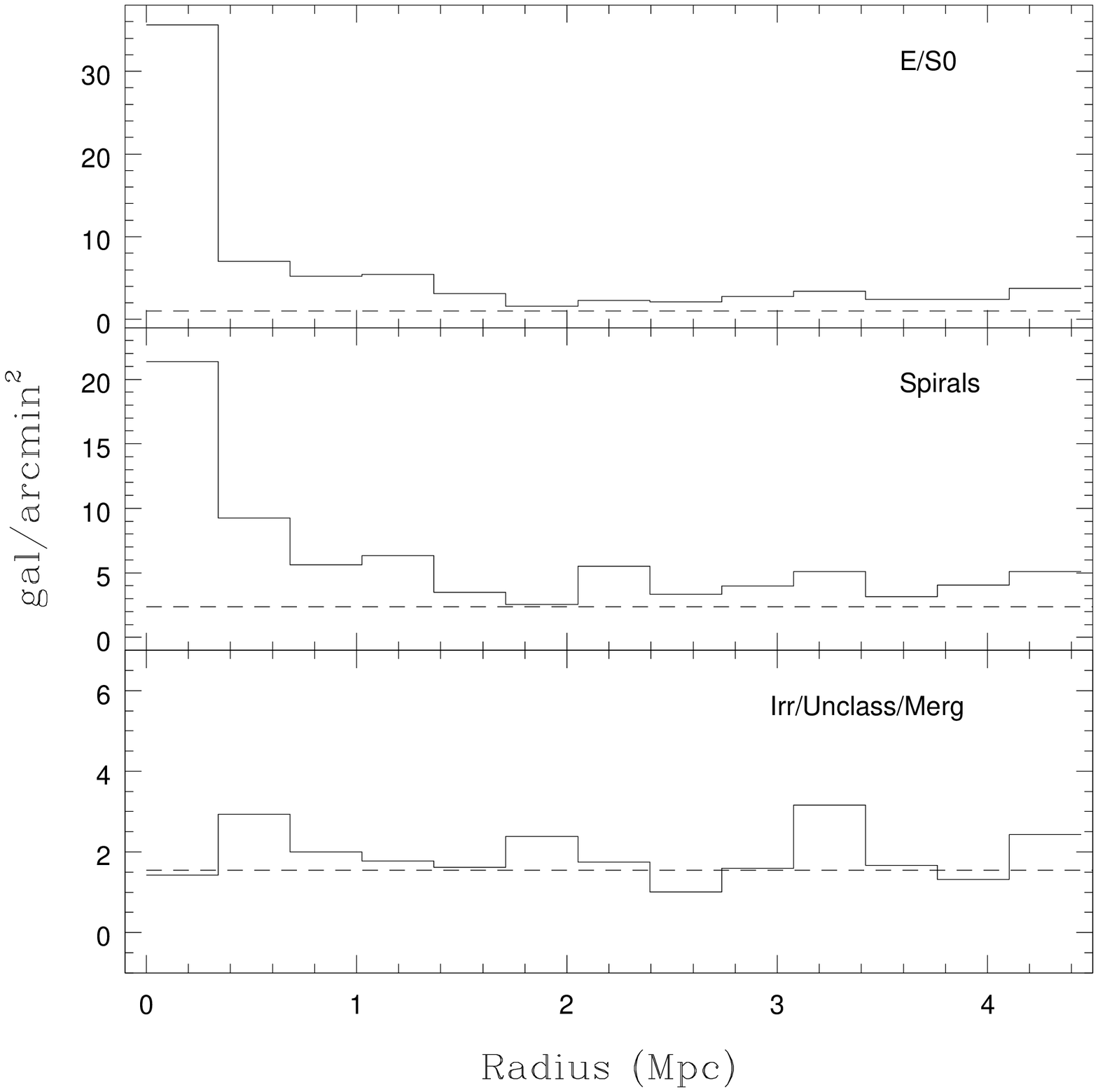}
\caption{{\em Left:} Galaxy surface density to $I<22.5$ as a function
of cluster-centric radius. Open symbols represent individual WF chips,
the filled symbols represent the average of a entire WFPC2 pointing.
The dashed line is the average field density (as inferred from Abraham
et al.\ 1996b). {\em Right:} As above, after azimuthal averaging.}
\end{figure}

What does the extensive spectroscopic data tell us about the
distribution of confirmed members in Figure 2? As discussed, the
kinematic structure of 0024 is more complex than that of a
single relaxed system. In particular, the distribution of radial
velocities is double peaked, with a secondary peak blueshifted by
3000 kms$^{-1}$ with respect to the main peak (Czoske et al.\
2001b). If we assign membership based on the redshift limits
adopted in Dressler et al.\ (1999), there are currently 215
confirmed members within our WFPC2 survey, 115 of which are E/S0s.

\begin{figure}
\plottwo{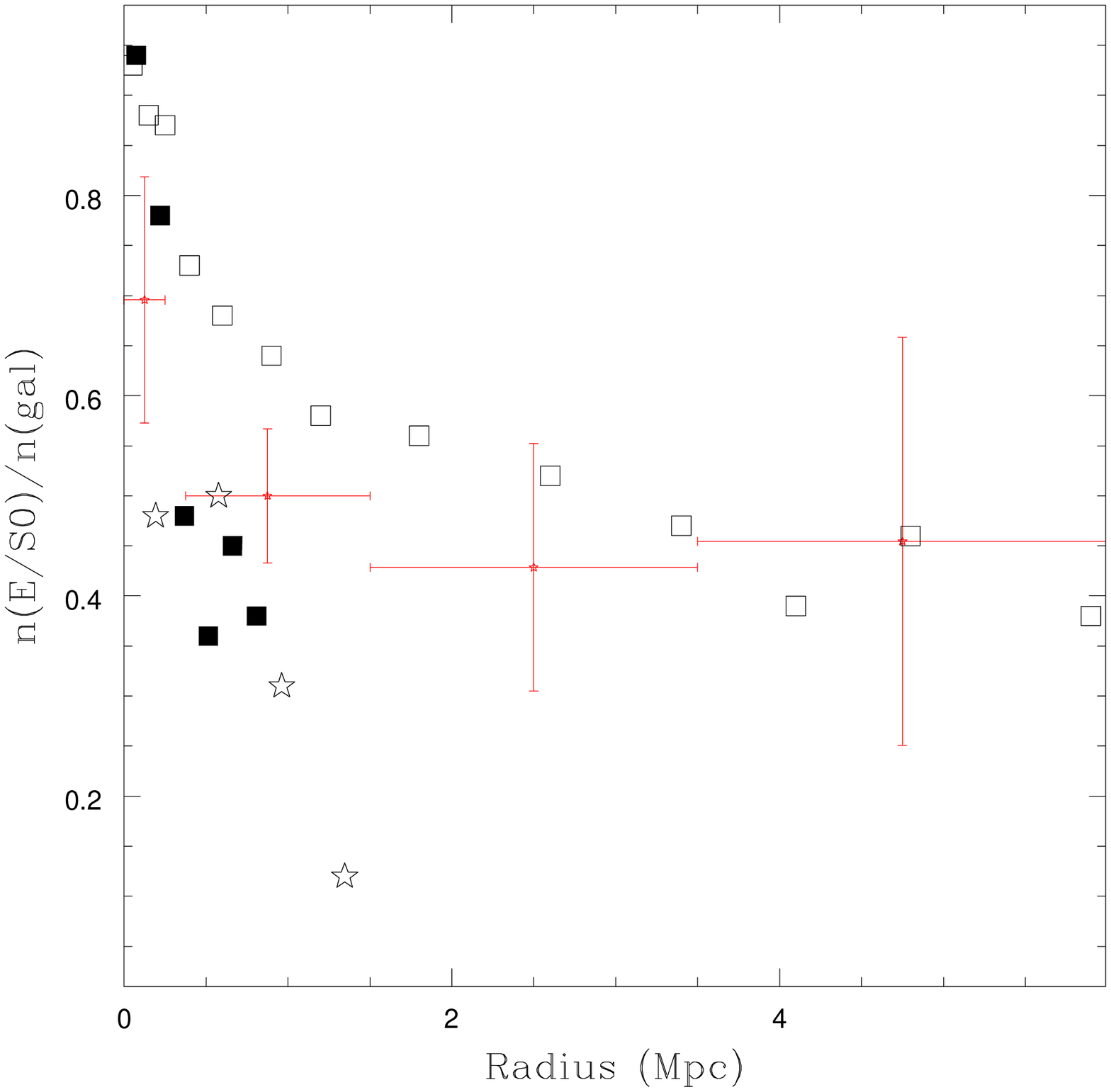}{treufig5.eps} 
\caption{{\em Left:} The fraction of E/S0s as a function of radius
(points with error bars) as compared to a local relation (open
squares, Whitmore, Gilmore \& Jones, 1993). Also shown are values
reported at $z=0.31$ (filled squares Couch et al.\ 1998) and at
$z=0.83$ (open stars; van Dokkum et al.\ 2000). {\em Right: } Shear
map (solid sticks) computed from the faint galaxy catalogue (small
ellipses). The cluster shear can be detected right to the periphery of
our HST data.}
\end{figure}

The left panel of Figure~3 shows the fraction of E/S0s in 0024
($I<21.1$ equivalent to the luminosity limit of Dressler 1980) as a
function of radius in comparison to a local relation and values
reported at intermediate redshift by Couch et al.\ ($z=0.31$) and van
Dokkum et al.\ (2000; $z=0.83$). Examining the trends across the
various clusters, the fraction of E/S0s in the cluster cores increases
with redshift, consistent with earlier findings by Dressler et al
(1997) and van Dokkum et al.\ (2000). However, perhaps surprisingly,
in the outer regions of 0024, the fraction of E/S0s remains
relatively high and consistent with the local value. Of course, the
uncertainties are large as the number of spectroscopic redshifts
beyond 2 Mpc remains small. If this result is borne out by enlarged
spectrosopic samples, one possible explanation is that the growth of
0024 is fuelled from merging substructures containing spheroidals
rather than solely via an abundant population of infalling isolated
field galaxies.  Alternatively, in the model proposed by Czoske et al
(2001b), some members of the core of the merging foreground component
will be scattered to large radii; in this case we would expect a
different dynamical distribution. 

\subsection{Weak lensing}

In classical studies of cluster dynamics, galaxies (or X-ray gas)
have been regarded as tracers of the gravitational potential under
a number of assumptions. Foremost it has been assumed that they
are not biased in any respect with respect to the dominant dark
matter potential. A significant advantage of studying in detail a
cluster at $z\simeq$0.4 is that it lies $\simeq$half-way to the
faint background population reached with a 2-orbit WFPC2 exposure.

It is possible, therefore, to contemplate combining the galaxy
dynamical data with the dark matter distribution as revealed by the
weak lensing induced by the cluster on the background population. As
an illustration of where we have reached with the recently arrived
data, Figures 3 and 4 show the first results arising from this
important additional physical diagnostic.

The right panel of Figure 3 shows the weak shear as a function of
position using a simple interpolation scheme to handle the
sparse-sampling. Remarkably, the shear is detected right to the
periphery of the HST mosaic where its value is typically
$\sim3\%$. Interestingly there is some discrepancy between the centre
of the shear pattern and that of the visible light, the
multiply-imaged arcs or the X-ray gas. This may be an artefact of the
preliminary reduction or it could represent contamination from the
foreground component. In the context of the model proposed by Czoske
et al.\ (2001b) this seems unlikely.

\begin{figure}
\plottwo{treufig6.eps}{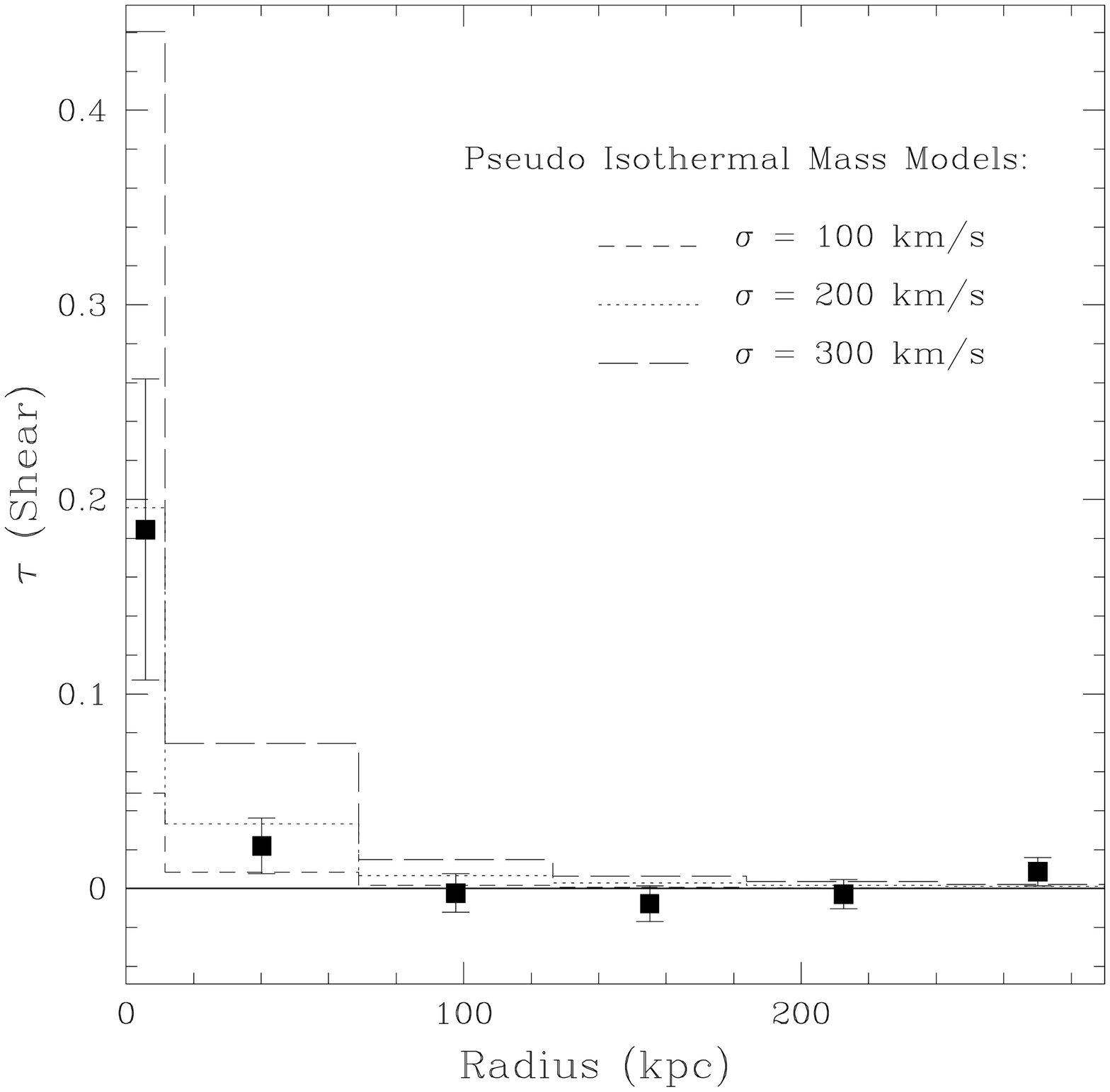}
\caption{{\em Left:} Mean shear as a function of cluster radius. An
extrapolation of the strong lensing model by Smail et al.\ (1996) is
shown, for different values of the average redshift of the background
objects ($1''$= $5.7$ kpc).  {\em Right: } Average galaxy-galaxy
lensing tangential shear $\tau$ versus radial distance from cluster
(lens) galaxy. Histograms represent the expected shear for
pseudo-isothermal mass distributions, assuming a core radius
$r_c=0.15$ kpc, truncation radius $r_t=30$ kpc and a source redshift
distribution peaked at $z_s=$1}
\end{figure}

\subsection{Galaxy-galaxy lensing}

A final application of this rich dataset is in determining the halo
properties of individual galaxies as probed by weak lensing induced by
individual cluster members. In dense environments, dark matter halos
of galaxies are likely to be truncated or stripped by environmental
effects (e.g. Moore et al.\ 1996). Previous studies in the cluster AC
114 at $z=0.31$ (Natarajan et al.\ 1998) suggested an apparent halo
cutoff for cluster members not seen in the equivalent field
population. While comprehensive measures of galaxy-galaxy lensing by
field galaxies have been carried out recently (e.g. McKay et al.\
2001), in clusters such studies have been limited by small sample
sizes and sample inhomogeneities.

The 0024 dataset can, in principle, probe halo properties in a
statistical sense from the high density core to the low density field
thereby securing a direct an unbiased test of the result claimed by
Natarajan et al. A preliminary analysis of the 38 WFPC2 pointings
yielded 137 cluster galaxies (71 with confirmed redshifts, 66 selected
on the color-magnitude diagram) and a carefully constructed background
galaxy sample of $\approx11,000$ objects. The shear obtained by direct
averaging (Natarajan \& Kneib 1997) is plotted in Fig.~4.

A shear detection of $\tau$ = 0.18 $\pm$ 0.08 (1 $\sigma$) is observed
in the innermost $2''$. Despite significant statistical uncertainties,
the observed value is consistent with that expected for an isothermal
mass distribution with $\sigma$ $\sim$ 200 km s$^{-1}$ (see caption
for detail). Compared to Natarajan et al.\ (1998) who detected $\tau$
= 0.16$^{+0.12}_{-0.13} $ (1 $\sigma$) in the innermost $1.5''$, the
galaxy-galaxy lensing signal is more clearly seen and to further
radial distance (see plot). Although still a weak signal, the improved
selection criteria and larger sample of spectroscopically confirmed
E/S0 galaxies are major advances over the AC114 study. The signal will
improve further when data from the Morphs field is used. Employing a
detailed cluster mass model which incorporates strong lensing
features, a maximum likelihood method (Natarajan \& Kneib 1997) will
be used to robustly extract halo parameters and look for radial trends
in halo mass and size.

\vfill


\begin{thebibliography}{}{
\bibitem{ACNOC} Abraham R.~G. et al.\ 1996a, \apj, 471, 694
\bibitem{AMDS} Abraham R.~G. et al.\ 1996b, \apjs, 107, 1
\bibitem{B00} Balogh M., et al.\ 2000, \apj, 540, 113
\bibitem{B00} Broadhurst T., et al. 2000, 534, L15
\bibitem{} Colless M. \& Dunn A. \ 1996, ApJ, 458, 435
\bibitem{C01a} Czoske O., et al. 2001a, A\&A, 372, 391
\bibitem{C01b} Czoske O., et al. 2001b, A\&A, submitted, astro-ph/0111118
\bibitem{dbK} de Blok et al. 2001, \apj, 552,L23
\bibitem{D80} Dressler A. 1980, \apj, 236, 351
\bibitem{D97} Dressler A. et al.\ 1997, \apj, 490, 577
\bibitem{D99} Dressler A. et al.\ 1999, \apjs, 122, 51
\bibitem{} Driver, S.P. et al.\ 1995, ApJ, 449, L23
\bibitem{} Kent, S. \& Gunn, J.E. \ 1982, AJ, 87, 945
\bibitem{K01} Kodama T., et al.\ 2001, \apj, 562, L9
\bibitem{Mc} McKay T.A. et al.\, 2001, submitted to \apj,
astro-ph/0108013
\bibitem{Moore} Moore B. et al.\ 1996, Nature, 379, 613
\bibitem{NK} Natarajan P. \& Kneib J.-P 1997, MNRAS, 287, 833.
\bibitem{P99} Poggianti B., et al.\ 1999, \apj, 518, 576
\bibitem{MORPHS} Smail I.~R. et al.\ 1997, \apjs, 479, 70
\bibitem{vD98} van Dokkum P.~G. et al.\ 2000, \apj, 541, 95
}
\end{thebibliography}
\end{document}